# High molecular gas fractions in normal massive star forming galaxies in the young Universe

L.J.Tacconi[1], R.Genzel[1,2], R.Neri[3], P.Cox[3], M.C.Cooper[4,5], K.Shapiro[6], A.Bolatto[7], N.Bouché[1], F.Bournaud[8], A.Burkert[9,10], F.Combes[11], J.Comerford[6], M.Davis[6], N.M. Förster Schreiber[1], S.Garcia-Burillo[12], J.Gracia-Carpio[1], D.Lutz[1], T.Naab[9], A.Omont[13], A.Shapley[14], A.Sternberg[15] & B.Weiner[4]

[1]*Max-Planck-Institut für extraterrestrische Physik (MPE), Giessenbachstr.1, 85748 Garching, Germany*

( linda@mpe.mpg.de, genzel@mpe.mpg.de )

[2]*Department of Physics, Le Conte Hall, University of California, 94720 Berkeley, USA*

[3]*IRAM, 300 Rue de la Piscine, 38406 St.Martin d'Heres, Grenoble, France*

[4]*Steward Observatory, 933 N. Cherry Ave., University of Arizona, Tucson AZ 85721-0065, USA*

[5]*Spitzer Fellow*

[6]*Department of Astronomy, Campbell Hall, University of California, Berkeley, 94720, USA*

[7]*Department of Astronomy, University of Maryland, College Park, MD 20742-2421, USA*

[8]*Service d'Astrophysique, DAPNIA, CEA/Saclay, F-91191 Gif-sur-Yvette Cedex, France*

[9]*Universitätssternwarte der Ludwig-Maximiliansuniversität, Scheinerstr. 1, D-81679 München, Germany*

[10]*Max-Planck-Fellow*

[11]*LERMA, Observatoire de Paris, 61 Av. de l'Observatoire, F-75 014 Paris, France*

[12]*Observatorio Astronómico Nacional-OAN, Apartado 1143, 28800 Alcalá de Henares- Madrid, Spain*

[13]*Institut d'Astrophysique de Paris, CNRS and Université Pierre and Marie Curie, 98 bis boulevard Arago, 75014 Paris, France*

[14]*Department of Physics & Astronomy, University of California, Los Angeles, CA 90095-1547, USA*

[15]*School of Physics and Astronomy, Tel Aviv University, Tel Aviv 69978, Israel*

**Stars form from cold molecular interstellar gas. Since this is relatively rare in the local Universe, galaxies like the Milky Way form only a few new stars per year. Typical massive galaxies in the distant Universe formed stars an order of magnitude more rapidly[1,2]. Unless star formation was significantly more efficient, this difference suggests that young galaxies were much more gas rich. Molecular gas observations in the distant Universe have so far been largely restricted to very luminous, rare objects, including mergers and quasars[3,4,5]. Here we report the results of a systematic survey of molecular gas in samples of typical massive star forming galaxies at $<z>\sim$1.2 and 2.3, when the Universe was 40% and 24% of its current age. Our measurements provide empirical evidence that distant star forming galaxies indeed were gas rich, and that the star formation efficiency is not strongly dependent on cosmic epoch. The average fraction of cold gas relative to total galaxy baryonic mass at z= 2.3 and z=1.2 is ~44% and 34%, three to ten times higher than in today's massive spiral galaxies[6]. The slow decrease between z~2 and 1 probably requires a mechanism of semi-continuous replenishment of fresh gas to the young galaxies.**

Direct observations of molecular gas in galaxies as a function of cosmic epoch are required to understand how galaxies have turned their gas into stars. To explore the evolution of cold gas fractions, we selected two samples of star forming galaxies (SFGs) spanning similar ranges in stellar mass and star formation rates: one at redshift z~1.2 ($t_0$~5.5 Gyr after the Big Bang) and the other at z~2.3 ($t_0$~3 Gyr). With recent improvements in instrumental sensitivity we can now sample the massive tail of the typical, or 'normal' star forming galaxies in this epoch (Supplementary Information, section 1). These 'main-sequence' galaxies are believed to have high (40-70%) duty cycles of star formation, and most are probably not 'starbursts' in a brief period of activity, such as major dissipative mergers[1,2]. Figure 1 displays the source-integrated spectra in the CO J=3–2 transition for 19 of the galaxies observed, 10 at z~2 and 9 at z~1. For 14 SFGs we have solid (>4σ) detections in both redshift ranges, and for the first time for z>2 SFGs. In 5 galaxies the emission is marginally or not detected, or may be continuum rather than line emission. Table 1 summarizes the observed and derived galaxy properties.

For EGS1305123 (z=1.12) and EGS1207881 (z=1.17) we also obtained high quality spatially resolved maps with FWHM resolutions of 0.65" and 1". Figure 2c shows the integrated CO 3-2 emission in EGS1305123, superposed on an optical image from the Hubble Space Telescope (HST). This map resolves for the first time the cold molecular gas distribution in a clearly non-merging high-z SFG. This system looks like a scaled-up version (in terms of star formation rate and gas mass) of normal z~0 gas rich disk galaxies. In the optical images EGS1305123 is a nearly face-on, large spiral disk. The clumpy CO emission extends over the entire disk, with a strong concentration of gas near the nucleus and innermost spiral arms (Figure 2 c). CO velocity channel maps (Figure 2, a & b) exhibit giant clumps with inferred gas masses of ~$5 \times 10^9$ $M_\odot$, intrinsic diameters <2-4 kpc, gas surface densities ≥500 $M_\odot pc^{-2}$ and velocity dispersions ~20 km/s. These clumps are correlated with, but typically separated by >1 kpc from the brightest nearby optical HII regions (similar to z~0 spirals, [12,13]). They probably represent loose conglomerates of several giant molecular clouds that are not yet resolved by our measurements, rather than a single, gravitationally stabilized, giant cloud, since the velocity dispersions are too low for their masses. These clumps are similar to but larger than conglomerates of molecular gas in z~0 spiral galaxies, which have masses of ~1-3x10$^7$ M$_\odot$, diameters of 500 pc, surface densities of ~100 M$_\odot$ pc$^{-2}$ and velocity dispersions of ~6-12 km/s [**12**]. The CO dynamics traces an ordered rotating disk pattern with maximum intrinsic rotation velocity $v_{d,max}$~200 km/s (Figure 2d-f). The inferred ratio of rotational velocity to line-of-sight velocity dispersion in the outer disk is $v_{d,max}/\sigma$~10, implying a fairly thin molecular gas disk but somewhat more turbulent than in local spirals[**12-13**]. The CO dynamics in EGS1207881 is also consistent with that of a large rotating disk.

The lower resolution observation galaxies, EGS13004291, EGS13017614, EGS13003805 and EGS12011767, all exhibit double-peaked line profiles with a spatial offset between the red and blue emission peaks, as expected for rotation in an extended disk (Supplementary Figure S2). This interpretation is consistent with the HST images (Supplementary Figure S2; images are available at http://tkserver.keck.hawaii.edu/egs/egsSurvey/egs_acsDownloads.php).

For BX610 at z=2.21 we detect a velocity gradient (Figure S2), which is consistent with the well defined rotation pattern in Hα line emission tracing ionized gas [**14**]. No published high resolution imaging or spectroscopy is available for the other z~2 galaxies in our sample. Based on Hα kinematics of 62 z~2 SFGs[**15**], we might expect between 1/3 and 2/3 of our massive z~2 SFGs to be rotating disks, similar to but probably more turbulent ($v_d/\sigma$~2-6) than the z~1 AEGIS galaxies. Future sub-arcsecond CO measurements will be necessary to reveal the molecular gas kinematics in these systems.

Our spatially resolved observations of EGS1305123 and the flux ratios of different CO lines in another z~1.5 SFG[16-17] are consistent with the CO emission in these systems arising in giant molecular cloud systems of temperature ~10-25 K and mean gas densities of $<n(H_2)>$~$10^2$ cm$^{-3}$, comparable to the Milky Way and z~0 SFGs[18-20]. In normal z~0 SFGs the CO line luminosity $L'_{CO}$ (K km/s pc$^2$) is proportional to the total cold (molecular hydrogen, plus helium) gas mass $M_{mol-gas}$. A similar 'Galactic' conversion factor appears to be also justified for the z~1-2 SFGs considered here, given the similar extended structure with large molecular cloud complexes and the comparable gas and star formation surface densities and near-solar metallicities of the z~1-2 SFGs and extragalactic the star forming clouds studied in references [**18-20**]. If anything a Galactic conversion factor may underestimate, rather than overestimate the total cold molecular gas masses (Supplementary Information, section 3)[20]. Figure 3 (a-c) shows the gas fractions derived in this way from Table 1 for all 19 z~1-2 SFGs (using 3$\sigma$ upper limits for the non-detections), and four SFGs from the literature[16,21-22].

The molecular gas fractions, defined as the ratio of gas mass to the sum of gas and stellar mass range broadly from 0.2 to 0.8, with an average of $<f_{mol-gas}>$~0.44 (Figure 3). SFGs at z=1-2 are three to ten times more gas-rich than z~0 SFGs with log$M_*$~10.5-11 ($<f_{mol-gas}$(spirals)$>$~0.04-0.1)[6, 23-24]. Our survey thus provides direct and statistically significant (~7-10 σ in the uncertainty of the mean, for a constant CO-$H_2$ conversion factor) empirical evidence for the long-standing expectation that high-z SFGs are much more gas rich than z~0 galaxies[25-27,8,14-17]. Within the uncertainties, the gas fractions do not depend much on galaxy selection method, star formation rate, or mass (Supplementary Information, sections 1 & 3). There appears to be a marginally significant trend (2.8σ in the uncertainty of the mean) that the z~2 SFGs are slightly more gas rich than those of similar mass at z=1, with average values of <$f_{mol-gas}$> ~0.44 and 0.34, respectively, broadly consistent with theoretical expectations (see below).

We caution that our sample is still limited and probes galaxies at the massive tail of the SFG populations in both redshift ranges (Figure S1). While the relative trends appear robust, the uncertainty in the derived absolute value of the gas fraction in each galaxy is substantial because of the combined uncertainties in the CO-$H_2$ conversion factor (Supplementary Information, section 3) and stellar masses, each at least ±50%. Larger samples will be needed to confirm the tantalizing redshift trend in Figure 3 b.

The comparably high gas fractions at z~2 and z~1 (separated by a cosmic time of ~2.5 Gyrs) are impressive, especially considering that the halo masses of most galaxies probably exceed $10^{12}$ $M_\odot$. These masses are close to or above the 'quenching mass' $M_t$~$5 \times 10^{11}$ $M_\odot$ at which the gas becomes hot and accretion inefficient[26-29]. The gas exhaustion time scales of our sample galaxies, $t_{exhaust}=M_{mol-gas}$/SFR, are ~0.9 Gyr (dispersion ±0.6 Gyr) for the z~1-2 SFGs, assuming that star formation rate continues at the current rate. These time scales are significantly shorter than the cosmic interval between z~2 and z~1, suggesting that either some replenishment is required during this epoch, or that the two samples have experienced different accretion and evolution histories. Our finding is consistent with simulations predicting that rapid accretion of cold gas can exist above $M_t$ and supplies the growing galaxies at high-z semi-continuously with fresh baryonic gas[25-27]. Our observations require that efficient feeding must continue in at least some massive galaxies to z~1 (Supplementary Information, section 4). Given their stellar masses and star formation rates it is possible that some of the z~1.2 galaxies are descendents of the types of galaxies we sample at z~2.3. On-going work on hydro-dynamical and semi-analytical simulations, including baryonic gas physics, star formation and feedback, predict average gas fractions similar to or somewhat lower than the observations in Figure 3 (25-45% at z~2, 10-40% at z~1-1.5, Davé et al. in preparation, Guo & White in preparation, Ocvirk et al. in preparation), all of which predict semi-continuous re-supply of the evolving galaxies with fresh gas from their surrounding cosmic web. This agreement between theory and observations is encouraging in terms of an overall emerging picture of galaxy formation.

A quantitative analysis of the data in Table 1 demonstrates that the large star formation rates at z~1-2 [1,2] are the consequence of the large molecular gas reservoirs and not of a greater star formation efficiency than at z~0. To within the uncertainties the so called 'Kennicutt-Schmidt' relation[30] between star formation rate and gas surface densities appears to be independent of redshift (Tacconi et al., in preparation).

Acknowledgments. We would like to thank Bernard Lazareff and the IRAM staff for their superb work in developing the new PdBI receiver systems, which made these technically difficult observations feasible. We are grateful to Jeremy Blaizot, Leo-Michel Dansac, Romeel Davé, Dušan Kereŝ, Pierre Ocvirk, Christophe Pichon and Romain Teyssier for communicating unpublished results of their simulations and for interesting discussions. This work is based on observations carried out with the IRAM Plateau de Bure Interferometer. IRAM is supported by INSU/CNRS (France), MPG (Germany) and IGN (Spain).

The authors have no competing financial interests.

Correspondence and requests for materials should be addressed to L. Tacconi (linda@mpe.mpg.de).

All authors have contributed extensively to this manuscript.

# Table 1. Properties of High-z Star Forming Galaxies[0]

| source[1] | z | $v_d$[2] | $R_{1/2}$[3] | SFR[4] | $F_{CO\,3-2}$[5] | $L_{CO\,3-2}$[6] | $M_{mol-gas}$[7] | $M_*$[8] | $f_{gas}$[9] |
|---|---|---|---|---|---|---|---|---|---|
| | | Km/s | kpc | $M_\odot$/yr | Jy km/s | K km/s pc$^2$ | $M_\odot$ | $M_\odot$ | |
| E13004291 | 1.20 | 300 | 7.2 | 172(86) | 3.7(0.15) | 3.2(0.13)e10 | 2.8(0.11)e11 | 3.3(1.3)e11 | 0.46(0.19) |
| E12007881 | 1.17 | 180 | 8.7 | 91(46) | 1.15(0.06) | 9.4(0.49)e9 | 8.3(0.43)e10 | 1.6(0.64)e11 | 0.34(0.14) |
| E13017614 | 1.18 | 270 | 6.6 | 74(37) | 1.25(0.10) | 1.1(0.08)e10 | 9.3(0.74)e10 | 1.1(0.42)e11 | 0.47(0.19) |
| E13035123 | 1.12 | 205 | 9.0 | 126(63) | 1.9(0.05) | 1.4(0.04)e10 | 1.3(0.03)e11 | 3.4(1.4)e11 | 0.27(0.11) |
| E13004661 | 1.19 | 180 | 6.6 | 82(41) | 0.32(0.06) | 2.8(0.52)e9 | 2.4(0.46)e10 | 3.0(1.2)e10 | 0.45(0.20) |
| E13003805 | 1.23 | 200 | 6.0 | 128(64) | 2.25(0.15) | 2.0(0.14)e10 | 1.8(0.12)e11 | 2.1(0.84)e11 | 0.46(0.19) |
| E12011767 | 1.28 | 80 | 7.0 | 47(24) | 0.30(0.06) | 2.9(0.59)e9 | 2.6(0.52)e10 | 1.2(0.48)e11 | 0.18(0.08) |
| E12012083 | 1.12 | 110 | 4.6 | 103(52) | <0.13(0.04) | <9.9(3.3)e8 | <8.7(2.9)e9 | 5.2(2.1)e10 | <0.14(0.07) |
| E13011439 | 1.10 | 94 | 4.6 | 90(45) | 0.70(0.15) | 5.1(1.1)e9 | 4.5(0.97)e10 | 1.3(0.5)e11 | 0.26(0.12) |
| BX 1439 | 2.19 | 265 | 8.0 | 97(43) | 0.23(0.08) | 6.6(2.2)e9 | 5.9(1.9)e10 | 5.7(2.3)e10 | 0.51(0.26) |
| BX 599 | 2.33 | 265 | 2.8 | 127(50) | 0.60(0.1) | 1.8(0.3)e10 | 1.6(0.26)e11 | 5.7(2.3)e10 | 0.73(0.32) |
| BX 663 | 2.43 | 256 | 5.5 | 131(49) | <0.18(0.06) | <5.7(1.9)e9 | <5.1(1.7)e10 | 6.9(2.8)e10 | <0.42(0.22) |
| MD 69 | 2.29 | 217 | 9.4 | 141(75) | 0.42(0.06) | 1.2(0.17)e10 | 1.1(0.15)e11 | 1.9(0.74)e11 | 0.36(0.15) |
| MD 94 | 2.34 | 217 | 9.6 | 382(171) | 2.0(0.3) | 6.0(0.9)e10 | 5.3(0.79)e11 | 1.5(0.61)e11 | 0.78(0.33) |
| MD 174 | 2.34 | 240 | 3.6 | 117(57) | 0.60(0.08) | 1.8(0.24)e10 | 1.6(0.21)e11 | 2.4(0.94)e11 | 0.4(0.17) |
| BX 691 | 2.19 | 238 | 6.7 | 62(27) | 0.15(0.05) | 4.0(1.2)e9 | 3.5(1.1)e10 | 7.6(3.0)e10 | 0.32(0.16) |
| BX 389 | 2.17 | 259 | 4.2 | 235(86) | <0.15(0.05) | <3.8(1.3)e9 | <3.3(1.1)e10 | 6.9(2.8)e10 | <0.33(0.17) |
| BX 442 | 2.18 | 238 | 6.7 | 92(45) | 0.43(0.08) | 1.1(0.21)e10 | 1.0(0.19)e11 | 1.5(0.59)e11 | 0.41(0.18) |
| BX 610 | 2.21 | 324 | 4.6 | 141(63) | 0.95(0.08) | 2.6(0.22)e10 | 2.3(0.19)e11 | 1.7(0.68)e11 | 0.57(0.23) |

[0] for details see Supplementary Information, sections 1,2,3

[1] Sources designated with 'E' are z~1 AEGIS galaxies, 'BX' and 'MD' for z~2 BX galaxies

[2] maximum intrinsic rotation velocity

[3] half-light radius

[4] extinction corrected star formation rates (and 1σ rms uncertainties) from a combination of UV/optical continuum, Hα and 24μm continuum, adopting a Chabrier[10] initial stellar mass function

[5] source and line integrated CO 3-2 flux with 1σ rms uncertainties in parentheses. Upper limits are 3σ rms.

[6] $L_{CO\ 3-2}=3.25\times10^{13}\ F_{CO\ 3-2}$(Jy km/s) $(D_L/Gpc)^2(1+z)^{-3}(\nu_{3-2,obs}/GHz)^{-2}$ (K km/s pc$^2$), where $D_L$ is the luminosity distance of the source, and $\nu_{3-2,obs}$ is the observed line frequency of the 3-2 line in GHz.

[7] total $H_2$ + He mass in cold gas (=1.36 times the $H_2$ mass), assuming $X=N(H_2)/I(CO)=2\times10^{20}$ cm$^{-2}$/(K km/s), or $\alpha=M(H_2)/L_{CO\ 1-0}=3.2$, determined from CO 3-2 luminosity and a correction I(CO 1-0)/I(CO 3-2)=2. The 1$\sigma$ rms uncertainties are in parentheses, and upper limits are 3$\sigma$ rms.

[8] stellar mass (and 1$\sigma$ rms uncertainties) determined from population synthesis modelling to the rest-frame UV- to infrared spectral energy distribution, assuming a Chabrier[10] initial stellar mass function

[9] $f_{mol-gas}=M_{mol-gas}/(M_{mol-gas}+M_*)$, upper limits are 3$\sigma$ rms

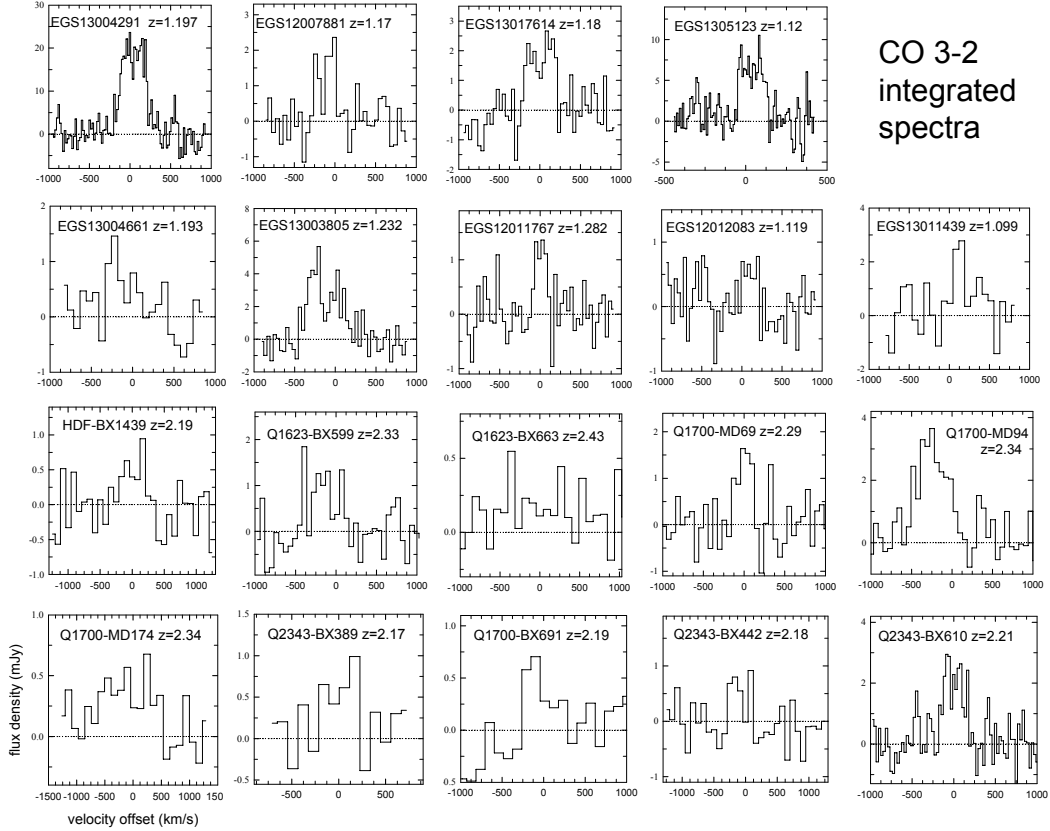

Figure 1. Integrated CO spectra. The 0.87mm CO J=3-2 rotational line spectra in typical SFGs at z>1. The z~1.2 SFGs were selected from the 'DEEP2/AEGIS' survey[7] by taking SFGs with a stellar mass $M_*>3\times10^{10}$ $M_\odot$ and a star formation rate SFR$\geq$40 $M_\odot$yr$^{-1}$ (Supplementary Information, section 1), and without major galaxy-galaxy interactions. The z~2.3 SFGs were selected from the H$\alpha$ survey of 'BX' UV-bright SFGs[8-9], with the same mass and star formation selection as at z~1.2. The <z>~1.2 (2.3) SFGs have mean stellar masses and star formation rates of <log( $M_*$ )>=11.11 (11.03 {$M_\odot$}) and <log(SFR)>=1.98 (2.13 {$M_\odot$yr$^{-1}$}) (Chabrier[10] initial stellar mass function) and sample the 'main sequence' of SFGs (see Figure S1).

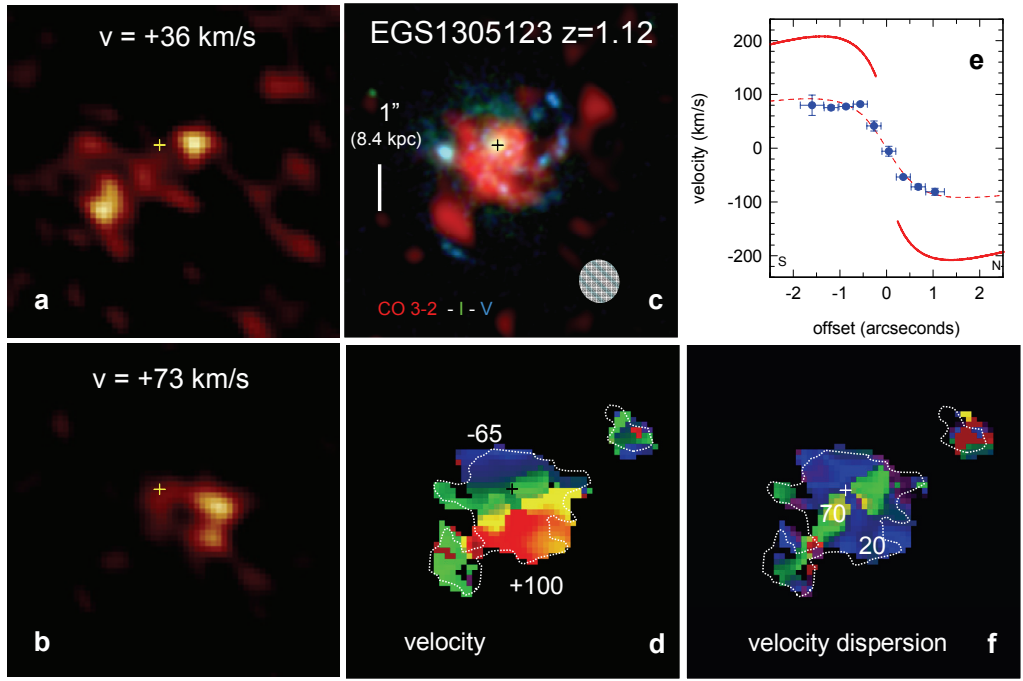

Figure 2. CO Maps in EGS1305123. High resolution (FWHM 0.6"x0.7") CO 3-2 maps and rotation curve in the z=1.12 AEGIS galaxy EGS1305123, obtained at ~2mm with the IRAM PdBI in A-configuration[11]. a & b: examples of two CO 3-2 maps in 9 km/s channels at +36 and +72 km/s; the rms in these channel maps is 0.3 mJy rms. The channel maps show several massive molecular clumps. The three brightest clumps visible on the + 73 km/s map have fluxes of 2.4, 2.1 and 1.45 mJy, or 8, 7 and 4.8 σ rms, respectively. The three bright clumps in the + 36 km/s map have fluxes of 1.7, 1.7 and 1.1 mJy, or 5.7, 5.7 and 3.7 σ rms, respectively. Typical gas masses in the clumps are ~5x10$^9$ M$_\odot$, intrinsic radii of <1-2 kpc, gas surface densities >300-700 M$_\odot$pc$^{-2}$ and velocity dispersions ~19 km/s. c: CO integrated line emission (red), I-band (green) and V-band (blue) HST ACS images of the source. The CO beam size is indicated by the hatched ellipse

and the cross marks the nucleus. d & f: peak velocity and velocity dispersion maps of the CO emission, obtained from Gaussian fits to the line emission in each spatial point of the map. The dotted white line shows the outline of the integrated emission and the cross marks the nucleus. e: peak CO velocity (and $1\sigma$ rms fitting uncertainty) along the major axis (p.a. 16° east of north) of the galaxy. The best fitting exponential disk model with radial scale length $R_d$=0.77" and dynamical mass of $2\times10^{11}$ $M_\odot$, for an adopted inclination of 27° is shown as a dotted red curve. The intrinsic rotation curve of this model as a function of radius is shown as a continuous red curve.

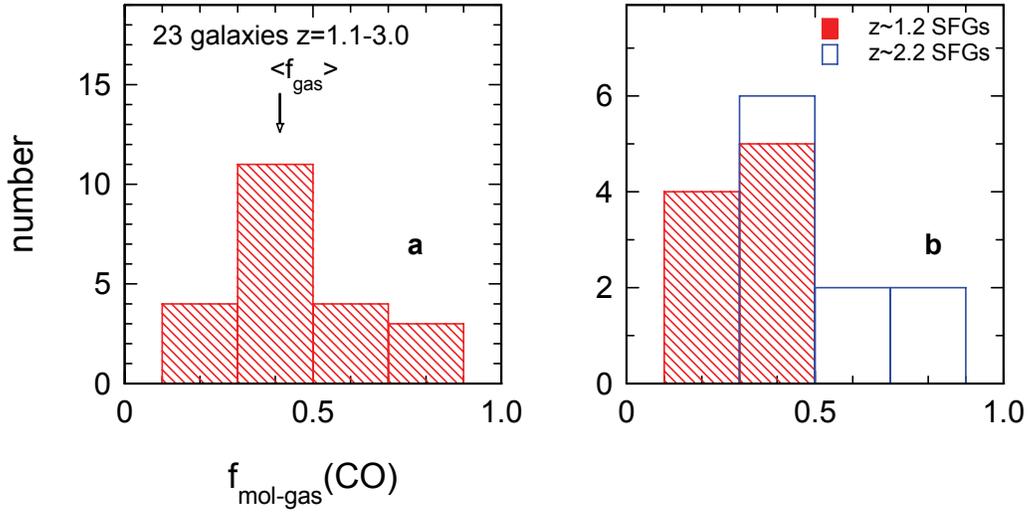

Figure 3. High molecular gas fractions in star forming galaxies at high-z. a) The distribution of molecular gas fractions, for all 23 SFGs with good stellar mass estimates from z~1-3.5, b) A comparison of the distribution of molecular gas fractions for the z~1 (red) and z~2 (blue) SFGs from this study. We define $f_{mol-gas}(2) = M_{gas}/(M_* + M_{gas})$. The molecular gas mass and fractions include a correction of 1.36 for helium.

# 1. Supplementary Information

## 1. Source selection

The high-z star forming galaxies selected in this paper were drawn from the AEGIS[1,7] and 'BX/BM'[8,9,31] galaxy surveys. The All-Wavelength Extended Groth Strip International Survey (AEGIS) provides deep imaging in all major wave bands from X-ray to radio (including Advanced Camera for Surveys (ACS) HST images) and optical spectroscopy (DEEP2/Keck) over a large area of sky (0.5 deg$^2$) with the aim of studying the panchromatic properties of galaxies over the last half of the Hubble time. The region studied is the Extended Groth Strip (EGS: RA=14$^h$17$^m$, Dec= 52$^0$30'). The AEGIS data provide the properties of a complete set of galaxies from 0.2≤z≤1.2 for the stellar mass range >10$^{10}$ M$_\odot$. Extinction corrected star formation rates are derived from a combination of Spitzer MIPS 24μm fluxes, GALEX UV fluxes and Hα/[OII] fluxes and use the Dale & Helou spectral energy distribution templates[1]. From the AEGIS data set we selected non-major merger (mass ratios of possible binary mergers >3:1) star forming galaxies (SFGs) with z~1.1-1.3, a stellar mass ≥3x10$^{10}$ M$_\odot$ and a star formation rate ≥40 M$_\odot$ yr$^{-1}$. Throughout we adopt a 'Chabrier/Kroupa' initial stellar mass function[10] and a ΛCDM cosmology with $H_0$=70 km/s and $\Omega_b$=0.046 and $\Omega_m$=0.28. The AEGIS galaxies presented in this paper were taken from the luminous end of the z~1.2 galaxy population but otherwise have star formation rates (SFRs) typical for (or slightly above) the 'main sequence' in the M$_*$- SFR plane (panel a of Figure S1). Our AEGIS sample thus samples the massive end of 'normal' star forming galaxies at z~1.2. Panels c and d of Figure S1 show that the derived molecular gas fraction at a given redshift does not depend on star formation rate or stellar mass. Together, these figures suggest that the z~1 AEGIS galaxies may have gas fractions representative of other z~1 'main sequence' SFGs and are therefore useful as a probe of the average molecular content of galaxies in this mass range at z~1.

Table 1 lists the star formation rates and stellar masses from AEGIS, as well as effective radii obtained from Sersic fitting to the ACS I-band images. In EGS13035123 and EGS1207881 we are also able to derive CO effective radii of 6.5 (±1.5) and 8 (±2) kpc, which are in reasonable agreement with the optical radii (9 and 8.7 kpc). The values of the maximum disk circular velocity $v_d$ quoted in Table 1 were computed from the velocity difference in the two line profile emission peaks, or from the velocity width and corrected for inclinations obtained from the minor to major axis ratios in the I-band images. In the case of EGS13035123 we used the model fit values to the CO 3-2 emission shown in Figure 2. Figure S2 shows ACS-HST V-band/I-band colour composites of all galaxies, along with CO 3-2 position velocity diagrams (where spatially resolved).

The z~2.3 SFGs were selected from the near-infrared long-slit spectroscopy sample of [**8**], which was drawn from the larger survey of [**9**], culled according to the so-called 'BX' criteria based on UGR colours (hereafter, simply BX sample). Our sub-sample was chosen from these surveys to cover the same stellar mass and star formation range as the z~1.2 AEGIS sample. Panel b of Figure S1 shows that our 'BX' sample, as our AEGIS sample, lies on the z~2 'main-sequence' and thus probes the high mass end of the entire

UV-/optically selected, 'normal' SFG population in this red-shift and mass range[2,31].
The majority of the BX galaxies also do not show evidence for undergoing major mergers [14,15,32] but BX663, MD94 and MD174 contain an active galactic nucleus based on the broad Hα lines observed[8]. Several of the BX galaxies detected in CO 3-2 were also part of the SINS integral field spectroscopy survey[14,15], where spatially resolved two dimensional kinematic data were obtained (for BX599, BX610, BX389, BX663). Figure S2 shows NICMOS-HST H-band images of BX610 from Förster Schreiber, Shapley and collaborators, along with a CO 3-2 position velocity diagram. BX610 clearly is a large, clumpy rotating disk. Likewise, the HST and SINS data for BX389 and BX663 show that these galaxies also are rotating disks (with a bright nuclear source or bulge in the case of BX663). For the rest of the BX galaxies observed here Hα data are in [8] but to our knowledge there are no HST images. The star formation rates listed in Table 1 are averages of the extinction corrected values obtained from stellar population synthesis fits to the rest-frame UV- to optical/near-infrared energy distributions and from Hα luminosities[8,15]. The Hα based instantaneous star formation rates use the conversions in (30) but we divide the 'Salpeter IMF' star formation rates by 1.7 to convert to a Chabrier/Kroupa[10] IMF, and apply the usual 'Calzetti'[33] correction to the continuum extinction values ($A_V$(nebular)=$A_V$(SED)/0.44, see discussion in [15]). Stellar masses are from spectral energy distribution fits[8,15], and $v_d$ and $R_{1/2}$ values are from the data in the same references with the methods discussed in [15]. The typical uncertainties of the derived quantities are dominated in most sources by systematic errors in stellar masses, star formation rates and $H_2$ masses, all of which are typically ±50%[8,15].

## 2. Observations and data analysis

The observations were carried out between June 2008 and June 2009 with the IRAM Plateau de Bure Millimetre Interferometer[11,34], which consists of six 15m-diameter telescopes. We observed the CO 3-2 rotational transition (rest frequency 345.998 GHz), which is shifted into the 2mm and 3mm bands for the z~1.2 and z~2.3 sources. For both bands our observations take advantage of the new generation, dual polarization receivers that deliver receiver temperatures of ~50 K single side band for both bands[11]. For source detections we used the 'C' and/or 'D' configurations of the instrument, resulting in ~5" and ~2" FWHM beam sizes for observations at 3 and 2 mm wavelength, respectively. For the high resolution observations of EGS13035123 and EGS1207881 we also used the 'B' and extended 'A' configurations (760 meter baseline), resulting in FWHM beam sizes of 0.6"x0.7" and 1" in these two sources. Weather conditions during the observations varied and data were weighted appropriate to their signal to noise ratio. Depending on the weather conditions and season, system temperatures (referred to above the atmosphere) were 100-200 K in both bands. Every 20 minutes we alternated source observations with a bright quasar calibrator within $15^0$ of the source. The absolute flux scale was calibrated on MWC349 ($S_{3mm}$=1.2 Jy). The spectral correlator was configured to cover 1 GHz per polarization. The source integration times were between 5 and 8 hours for the detections and a total of 20 hours for the highest resolution mapping in EGS13035123. The data were calibrated using the CLIC package of the IRAM GILDAS software system and further analyzed and mapped in the MAPPING environment of GILDAS. Final maps were cleaned with the CLARK version of CLEAN implemented in GILDAS. The absolute flux scale is better than ±20%.

## 3. Estimation of $H_2$ column densities and masses

Observations in giant molecular clouds (GMCs) of the Milky Way have established that the integrated line flux of $^{12}CO$ millimetre rotational lines can be used to infer cold (molecular) gas masses, despite the facts that the CO molecule only makes up a small fraction of the entire gas mass and that the lower rotational lines (1-0, 2-1, 3-2) are almost always very optically thick[18,19]. This is because the CO emission in the Milky Way and nearby normal galaxies comes from moderately dense (volume averaged densities $<n(H_2)> \sim 200$ cm$^{-3}$), self-gravitating GMCs. In this regime the ratio of $H_2$ column density to integrated CO flux I(CO) (I(CO)= $\int_{line} T_R(v) \, dv$), X, or the ratio of $H_2$ mass to CO luminosity $L'_{CO}$ ($L'_{CO}=\int_{source}\int_{line} T_R(v) \, dv \, dA$) can be expressed as

$$N(H_2)/I(CO) = X = c_1 \left(\frac{<n(H_2)>}{200 \text{ cm}^{-2}}\right)^{1/2} \left(\frac{T_R}{6 \text{ K}}\right)^{-1} \quad , \quad [\text{cm}^{-2}/(\text{K km s}^{-1})]$$

and

$$M_{H_2}/L'_{CO} = \alpha = c_2 \left(\frac{<n(H_2)>}{200 \text{ cm}^{-2}}\right)^{1/2} \left(\frac{T_R}{6 \text{ K}}\right)^{-1} \quad , \quad [M_\odot/(\text{K km s}^{-1}\text{pc}^2)] \; .$$

Here $T_R$ is the equivalent Rayleigh-Jeans brightness temperature of the (optically thick) CO line and $c_1$ and $c_2$ are appropriate numerical constants. In the 2.6mm CO (1-0) transition the typical gas temperature of Galactic GMCs is ~10-25 K. Several independent empirical techniques based on GeV γ-rays, optical extinction measurements, isotopomeric line ratios and excitation analysis have all shown that this 'virial' technique is appropriate and remarkably robust throughout the Milky Way[18,19,35,36]. The best empirical conversion factor is X~2x10$^{20}$ [cm$^{-2}$ (K km s$^{-1}$)$^{-1}$] and α=3.2 [M$_\odot$(K km/s pc$^2$)$^{-1}$ ]. To calculate the total amount of cold gas in the molecular phase, the H$_2$ masses have to be multiplied by a factor of 1.36 to account for interstellar helium. The virial approach can be shown to also apply to an ensemble of virialized clouds, instead of a single one, as long as the factor n(H$_2$)$^{1/2}$/T is constant throughout the system and the CO line is optically thick[19]. Bolatto et al. (reference [20]) have observed and discussed spatially resolved CO measurements of star forming clouds in Local Group, large and dwarf galaxies, with a range of metallicity from solar to SMC metallicity. They find that the CO conversion factor in these star forming cores, with gas surface densities and star formation surface densities very comparable to our z~1-2 SFGs (for Table 1: <Σ$_{H2}$>~250 M$_\odot$pc$^{-2}$, <Σ$_{SFR}$>=0.5 M$_\odot$ yr$^{-1}$ kpc$^{-2}$), is close to or perhaps somewhat above the Galactic value. In z~0 dwarf galaxies with sub-solar metallicity (probably applicable to most of the lower mass systems in Table 1), there is evidence for additional, extended and CO-faint molecular gas outside of the bright star forming clouds, so that the galaxy-wide CO to H$_2$ conversion factor in these systems is above the Galactic value (see discussion and references in [5] and [20]). As such, we believe that the Galactic conversion factor we have chosen for the z~1-2 SFGs is a conservative lower limit.

In reality the conversion factor is probably a function of parameters, including gas or star formation surface density and metallicity[5,37]. Future observations are required to establish this 'conversion function' empirically. However, the main points of this paper relating to the comparison of relative gas fractions in star forming galaxies at z~2.3, 1.2 and 0 are probably fairly robust, since we are comparing galaxies of comparable gas and star formation volume and surface density, masses and metallicities.

For the z~1-2 SFGs a 'Galactic' conversion factor ($\alpha$~3.2 $M_\odot$/(K km s$^{-1}$ pc$^2$)) is appropriate since the CO emission in these systems, as in z~0 disk galaxies, probably arises in virialized giant molecular cloud systems (GMCs) of temperature ~10-25 K and mean gas densities of $<n(H_2)> \sim 10^{2...3}$ cm$^{-3}$ [17]. However, to convert the CO 3-2 luminosity to an equivalent CO 1-0 luminosity we apply in Table 1 a correction factor of $R_{13}=L'_{CO\ 1-0}/L'_{CO\ 3-2}$~2 to correct for the lower Rayleigh-Jeans brightness temperature of the 3-2 transition relative to that of the 1-0 transition, caused by the Planck correction at low temperature and by the sub-thermal population in the upper level of the transition. This correction factor is motivated by CO 3-2/1-0 observations both in z~0 disks[38] and in the z~1.5 SFG BzK21000[17].

For the large column densities (and interstellar pressures) in the z>1 SFGs ($\Sigma_{gas}$>>10 $M_\odot$ pc$^{-2}$) most of the cold interstellar gas is probably in molecular form and the contribution of atomic hydrogen can be neglected[39].

## 4. Semi-Continuous Fuelling of Gas into Massive Galaxies at z>1

The molecular data presented here provide, for the first time, direct molecular gas masses for a number of z~1-2 massive galaxies. Coupled with the star formation rates, these masses yield gas exhaustion timescales of ~0.5-1 Gyr for these galaxies; if the star formation activity remains constant for longer than a fraction of a Gyr, then a semi-continuous and efficient replenishment of gas is needed.

Other recent observations, based mainly on the statistical properties of large samples, have shown that continuous gas inflows from the halos and continuous star formation history are required to explain the properties of high-redshift galaxies. The existence and relatively small scatter of the 'main sequence' has been interpreted, along with evidence based on the observed cosmic abundances of the 'main sequence' galaxies (relative to expectations from the halo masses they are embedded in), and the ratio of star forming to non-star forming galaxies, that galaxies near the main sequence have a large (40-70%) duty cycle of star formation[1,2]. This means that most form stars regularly at a near constant rate and cannot be 'starbursts' during a short period of high activity, such as dissipative major mergers.[1,2,31,40] (see also above). Analytic arguments based on the observed metallicities of the z~2 population show that these SFGs are likely powered by a continuous inflow rate equal to twice the SFR (coupled to an outflow rate equal to the SFR)[41]. Further support that this inflow be somewhat smooth with embedded clumps ("cold flow" or a series of minor mergers, not a major merger) comes from the kinematics of z~2 'main sequence' galaxies, of which the majority show no signs of recent major merger-like interactions despite their high SFRs[14,32,42].

Numerical simulations have also added increased support to the semi-continuous replenishment scenario. Analysis of the accretion histories of dark matter haloes in the Millennium simulation indicates that only a small fraction of galaxies with the masses and SFRs of z~2 SFGs would be expected to have undergone a recent major merger[43], which would be able to explain the observed star formation rates and perhaps in some cases also the kinematic properties of the clumpy rotating disks[44]. Instead, their high SFRs are fuelled by accretion of diffuse matter small clumps (minor mergers) that are not associated with larger dark matter halos. The baryonic physics associated with this process have been demonstrated by e.g. (**27**), who show that filaments of cold gas can penetrate to the centre of a galaxy halo and provide molecular fuel in a smooth manner. Finally, (**43**) have shown that the observed number of z~2 SFGs constitute a significant fraction of the number of dark matter haloes in the respective mass range; this implies that these galaxies are not merely the "on" phase of a larger, quiescent population and therefore that the smooth accretion processes inferred for the observed galaxies are representative of the galaxy population in this mass range as a whole.

## 2. Supplementary Figures

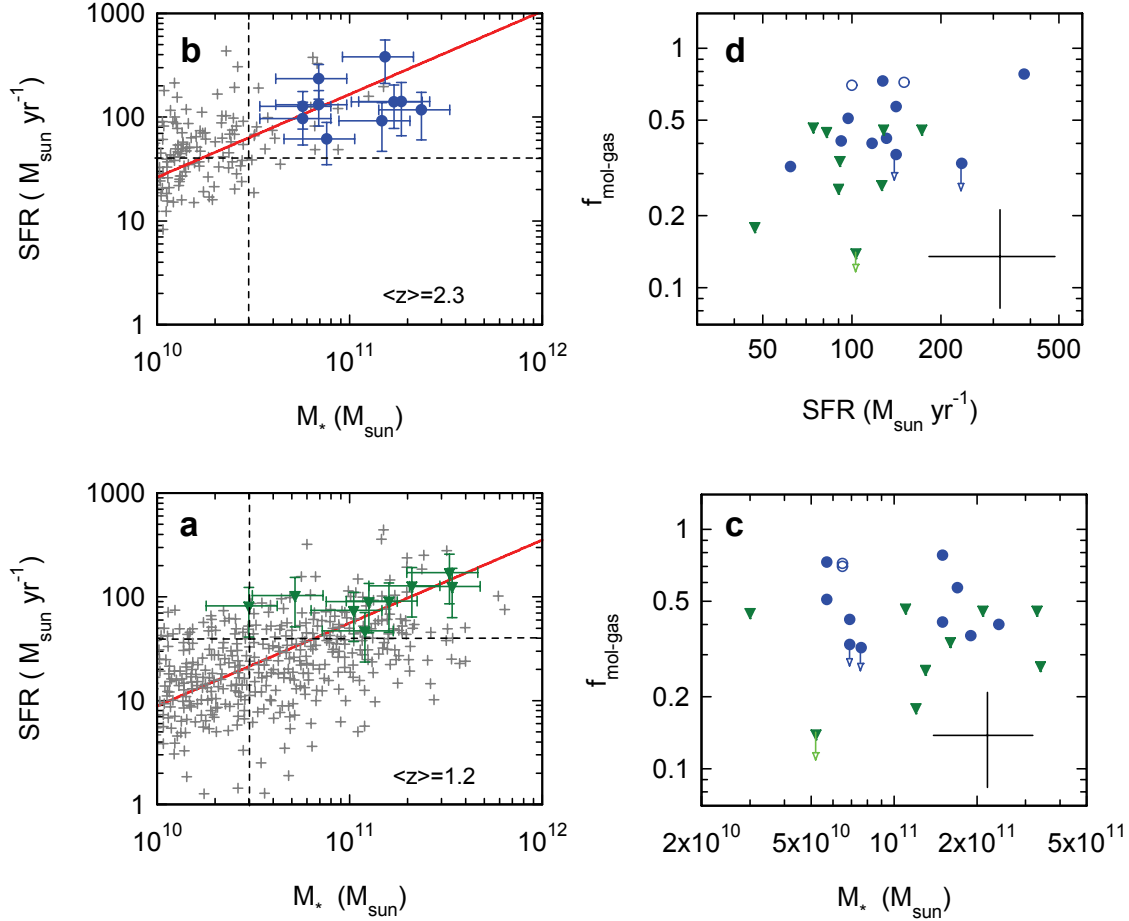

Fig. S1: Properties and parameter correlations of high-z SFGs with molecular gas masses. a and b: Location of the z~1.2 (a) and z~2.3 (b) SFGs in the stellar mass-star formation rate plane. The thick red line marks the location of the best fit to the $M_*$-SFR distribution at z~1.2 and 2.3 (the 'main-sequence line', **1,2**: SFR $(M_\odot yr^{-1})$=150$(M_*/10^{11} M_\odot)^{0.8}([1+z]/3.2)^{2.7}$, Bouché et al. in preparation). The grey crosses shows the SFR-$M_*$ data in (**1,2**) scaled to the same mean redshift as the CO observations with the $(1+z)^{2.7}$ dependence given above. The dashed vertical and horizontal lines mark the common matched selection criteria for both red-shift ranges ($M_* \geq 3\times10^{10}$ $M_\odot$, SFR$\geq$40 $M_\odot$ yr$^{-1}$). c

and d: Dependence of the molecular gas fractions in Table 1 on stellar mass and star formation rate. In addition to the z~1.2 and 2.3 SFGs we added two z~1.5 SFGs[16] (open blue circles). Large crosses denote the typical measurement uncertainties, which are dominated in most cases by systematic errors.

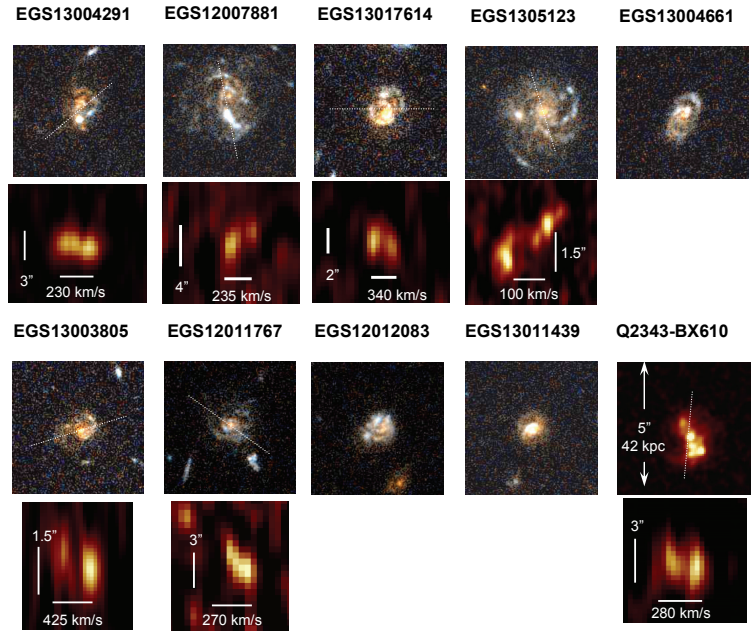

Figure S2. HST images and CO 3-2 position velocity diagrams for some of the program galaxies. For the nine z~1.2 AEGIS SFGs we show on top a V-band+I-band composite ACS image for each of the galaxies. Below we show, where derivable, a CO 3-2 position-velocity diagram along the direction denoted by a dotted line in the HST image. For the z~2.2 SFG BX610 we show a NICMOS H-band image from Förster Schreiber, Shapley and collaborators on top.

## 3. Supplementary Notes

## Extra references used in the Supplementary Information